\documentclass[fleqn,11pt]{wlscirep}
\usepackage[utf8]{inputenc}
\usepackage[T1]{fontenc}
\usepackage{siunitx}  
\usepackage[version=4]{mhchem}

\newcommand{\rp}{\mathbf{r}_{\perp}}
\newcommand{\abs}[1]{\left| #1 \right|}
\newcommand{\kp}{\mathbf{k}_\perp}
\DeclareSIUnit\Oe{Oe}

\title{AI-enabled Lorentz microscopy for quantitative imaging of nanoscale magnetic spin textures}

\author[1,2]{Arthur R. C. McCray}
\author[3]{Tao Zhou}
\author[4]{Saugat Kandel}
\author[1,5]{Amanda Petford-Long}
\author[4]{Mathew J. Cherukara}
\author[1,5,*]{Charudatta Phatak}
\affil[1]{Materials Science Division, Argonne National Laboratory, Lemont, IL, USA}
\affil[2]{Applied Physics Program, Northwestern University, Evanston, IL, USA}
\affil[3]{Center for Nanoscale Materials, Argonne National Laboratory, Lemont, IL, USA}
\affil[4]{Advanced Photon Source, Argonne National Laboratory, Lemont, IL, USA}
\affil[5]{Department of Materials Science and Engineering, Northwestern University, Evanston, IL, USA}

\affil[*]{cd@anl.gov}

\begin{abstract} 
The manipulation and control of nanoscale magnetic spin textures is of rising interest as they are potential foundational units in next-generation computing paradigms. Achieving this requires a quantitative understanding of the spin texture behavior under external stimuli using in situ experiments. Lorentz transmission electron microscopy (LTEM) enables real-space imaging of spin textures at the nanoscale, but quantitative characterization of in situ data is extremely challenging. Here, we present an AI-enabled phase-retrieval method based on integrating a generative deep image prior with an image formation forward model for LTEM. Our approach uses a single out-of-focus image for phase retrieval and achieves significantly higher accuracy and robustness to noise compared to existing methods. Furthermore, our method is capable of isolating sample heterogeneities from magnetic contrast, as shown by application to simulated and experimental data. This approach allows quantitative phase reconstruction of in situ data and can also enable near real-time quantitative magnetic imaging.   
\end{abstract}
\begin{document}

\flushbottom
\maketitle
\section*{Introduction} 
Magnetic spin textures on the micro- and nano-scale are of continued importance for both a fundamental understanding of their physical behavior and for potential applications in novel computing paradigms \cite{sharma2022, li2021b, wang2022b}. Topologically-protected spin textures, such as skyrmions, have been shown to be highly mobile and stable \cite{Fert2017, Everschor-Sitte2018}. This makes them appealing candidates for information carriers, and their non-trivial topology, when interfaced with other materials, gives rise to interesting physics including topological superconductivity and Majorana bound states \cite{Dahir2019, Mascot2021, Iwasaki2014}. Magnetic skyrmion-based neuromorphic and reservoir computing schemes have been developed, and these motivate future studies of how to control individual and collective skyrmion behavior under external stimuli such as temperature or magnetic and electric fields \cite{raab2022, pinna2020, msiska2023, zhang2020}. Materials that are being explored for such phenomena include ultra-thin multilayer structures such as \ce{Pt/Co/X} as well as van der Waals (vdW) ferromagnets such as \ce{Cr2Ge2Te6} (CGT), \ce{Fe3GeTe2}, and \ce{CrI3} \cite{wang2022a, gibertini2019, kurebayashi2022, burch2018}. For all of these materials, real-space imaging is critical for observing both the structure of individual magnetic spin textures and also how they respond to external stimuli.  

Lorentz transmission electron microscopy (LTEM) is a powerful technique for imaging magnetic spin textures in these materials and other thin films, as it allows the simultaneous observation of magnetic domains and microstructure \cite{phatak2016}. Furthermore, LTEM can be used to perform a wide range of in situ experiments that allow for imaging the behavior of spin textures under stimuli such as variable temperature, applied magnetic field, and electric current. \cite{phatak2016, ngo2016} LTEM can also be used to obtain quantitative information about the sample's magnetic induction, which is related to the sample magnetization, and is carried by the phase of the electron wave as described by the Aharonov-Bohm equation \cite{Aharonov1961}. In the Aharonov-Bohm framework, the total electron phase shift is comprised of two components, $\phi = \phi_e + \phi_m$, where $\phi_e$ is the electrostatic component dependent on the sample material, thickness, and local electrostatic potential, and $\phi_m$ is dependent on the magnetic vector potential. If the total electron phase shift, $\phi$, can be separated to remove the electrostatic component, $\phi_e$, and isolate the magnetic phase shift, $\phi_m$, the gradient of $\phi_m$ can be used to calculate the in-plane component of the integrated magnetic induction, which is perpendicular to the electron beam direction. 

Phase retrieval is commonly achieved by solving the transport of intensity equation (TIE), which relates $\phi$ to a through-focal series (TFS) of images \cite{Teague1983, DeGraef2001}. The TIE approach is easy to implement but suffers from reduced spatial and phase resolution compared to other techniques that give higher resolution at the expense of additional experimental requirements \cite{McVitie2006, Koch2014}. For example, off-axis holography requires an electron biprism and the ability to obtain a reference electron wave, making it unsuitable for many samples \cite{Gabor1948, anada2019, iwasaki2023}, and 4D-STEM requires long measurement times and pixelated detectors \cite{Ophus2019}. Recently, a machine-learning-based approach was developed to apply direct automatic differentiation (AD) to a forward model, which has been shown to reconstruct $\phi$ more accurately than the TIE method when using an input TFS \cite{Zhou2021}. 

All of these phase reconstruction techniques require either multiple images, long acquisition times, or complicated experimental setups and are therefore unsuitable for in situ experiments that study the time evolution of magnetic spin textures. In many cases, a movie or a sequence of defocused LTEM images are the only data that can be collected. There exists a useful modification of the TIE, known as single image TIE (SITIE), that enables phase reconstruction for some samples \cite{Chess2017, paganin2002}. SITIE, however, is only accurate in the small defocus limit and assumes that all contrast is magnetic in origin. This makes it very susceptible to noise and to the influence of non-magnetic contrast such as diffraction contrast. In addition, real-world samples can often contain non-magnetic regions, including focused ion beam damage, ice that accumulates during cryogenic experiments, or surface contaminants. These all add contrast that will be incorrectly reconstructed as magnetic domains. Despite these drawbacks, SITIE is frequently used for mapping the integrated magnetic induction of thin films and 2D materials \cite{kim2020, McVitie2018, Mccray2021}. In almost all cases it is used qualitatively, as techniques for dealing with noise, including Tikhonov filtering or image pre-processing, can make the resulting reconstructions non-quantitative \cite{Mitome2010}.

In this work, we present the development and demonstration of Single Image Phase Reconstruction via Automatic Differentiation (SIPRAD): a method for reconstructing the magnetic phase shift, $\phi_m$, from a single defocused LTEM image. SIPRAD is shown to be quantitatively accurate and robust to high levels of noise across a wide defocus range. It is also uniquely able to isolate $\phi_m$ in samples with amplitude and $\phi_e$ variations that are present because of artifacts or sample heterogeneity. These characteristics make SIPRAD ideally suited for performing phase reconstructions from in situ experiments.

\section*{Results} 
\subsection*{Single image phase retrieval}
\begin{figure}[htb]
\centering
  \includegraphics[width=0.9 \linewidth]{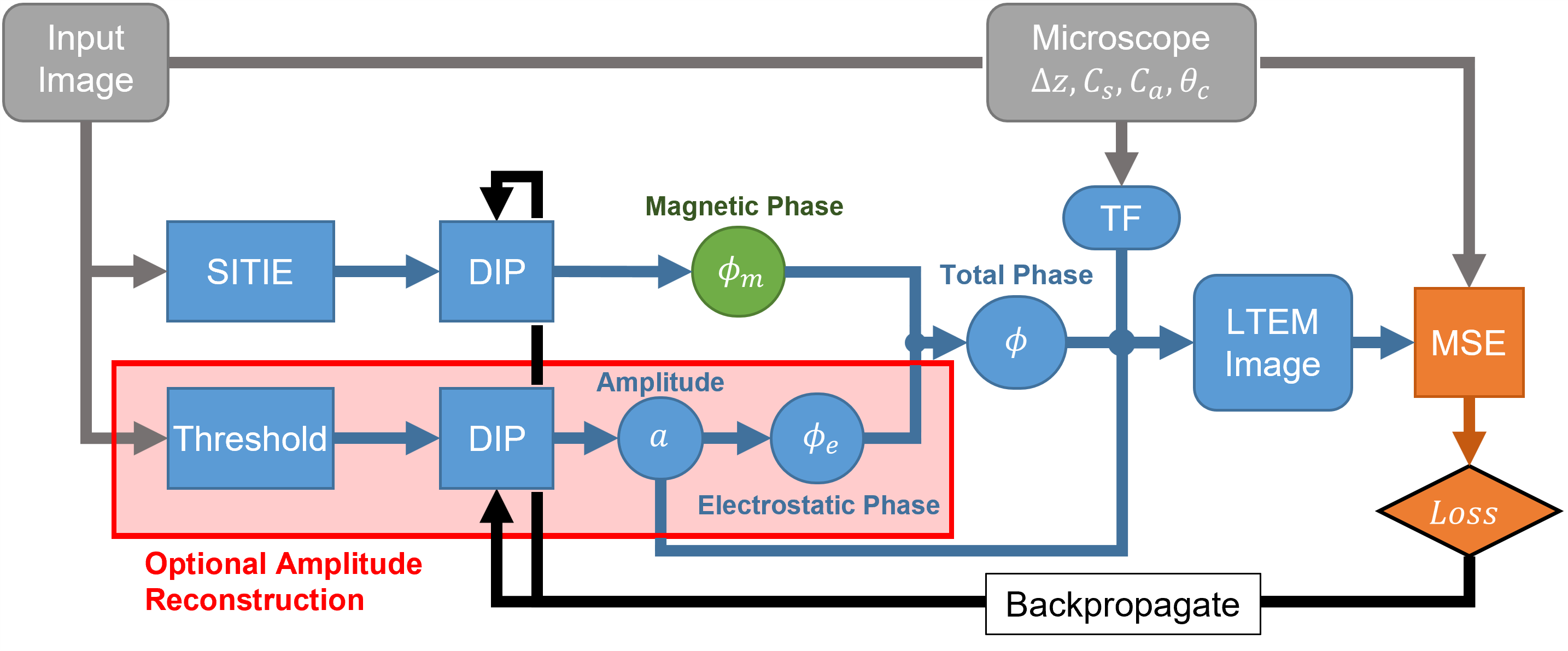}
  \caption{\textbf{Overview of the SIPRAD magnetic phase retrieval algorithm.} A single defocused LTEM image and microscope parameters are given as input, and a deep image prior (DIP) is trained to output $\phi_m$. If the amplitude reconstruction branch is activated, a second DIP will be trained to output an amplitude map which will also be scaled according to material parameters and used as the electrostatic phase shift $\phi_e$.}
  \label{fig:outline}
\end{figure}

Figure \ref{fig:outline} depicts a schematic of the SIPRAD algorithm that uses AD to reconstruct $\phi_m$. In previous work, AD was applied directly to an image formation model in order to reconstruct the total phase shift $\phi$ from a TFS of images \cite{Zhou2021}. With only a single, often noisy, image, applying this same technique leads the algorithm either to over-fit to noise or diverge. To avoid this, a key feature of our algorithm is that it uses up to two deep-image-priors (DIPs) in place of other regularization techniques. By doing so we are able to perform accurate phase reconstructions from a single noisy image. 

As described in Ulyanov et al., a DIP is a generative convolutional neural network (CNN) that is trained to create a desired image \cite{ulyanov2018}. The DIP relies upon the CNN architecture to generate images with lower patch-wise entropy, and DIPs have been successfully applied for tasks including image de-noising, segmentation, and tomographic reconstructions \cite{ulyanov2018, gandelsman2019, zhou2020, du2021}. We find that using a DIP both stabilizes the algorithm and provides superior noise robustness compared to other techniques such as total-variance regularization, with the additional benefit of eliminating user-optimized parameters. Therefore a key distinction in our algorithm from previous AD phase retrieval approaches is that, rather than attempting to learn the phase shift directly, a DIP is trained to generate the reconstructed phase shift.

We will begin by considering a case where the amplitude and $\phi_e$ are both uniform across a sample, such as for a uniformly thick flake of vdW material or amorphous film. The branch of the algorithm that reconstructs variations in the electron wave amplitude, which will be discussed further, is therefore turned off. In all cases the algorithm begins with an untrained DIP that is given an approximate phase input and which iteratively learns to output an improved $\phi_m$. In many previous applications of DIPs, random noise is used as the input \cite{ulyanov2018, gandelsman2019, zhou2020, du2021}. We have found that using an approximate phase shift, obtained with the SITIE method, as an input for the DIP, can allow the algorithm to converge more quickly. The output phase shift of the DIP, $\phi_m$, is used as the input of an image-formation forward model with microscope parameters set by the user to match experimental values. The mean-squared error (MSE) loss is calculated between this resulting simulated LTEM image and the input experimental image. Gradients are calculated by back-propagating the loss through the algorithm and are then used to update the weights of the DIP, iteratively improving the accuracy of its output phase. 

The image formation forward model begins by writing the electron exit wavefunction at the sample as $\psi \left( \rp \right) = a \left( \rp \right) e^{i \phi \left( \rp \right)}$, where $a \left( \rp \right)$ is the amplitude function the electron wave that depends on the sample shape function and composition, $\phi$ the total phase shift imparted by the sample, and $\rp$ is a radial vector perpendicular to the electron beam direction, which is taken to be $z$. This wavefunction is then convolved with the microscope transfer function, which consists of an aperture function, phase transfer function, and damping envelope. This linear image formation model has been shown to work well for LTEM, as the deflection angle due to the Lorentz force is very small compared to that of the diffracted beams \cite{Zhou2021, Mccray2021, Chess2017}. Additional details regarding the forward model are given in the ``Methods'' section. 

Our goal is to reconstruct $\phi_m$ and, for samples with uniform amplitude and uniform $\phi_e$, a single DIP is sufficient because $\phi \sim \phi_m$ except for a uniform offset. The SIPRAD method also provides a way to simultaneously reconstruct $\phi_m$ and separate it from the non-uniform amplitude function and $\phi_e$ that arise from sample heterogeneity and contamination. For the examples in this work, we assume that the origin of a non-uniform electron wave amplitude are particles on the sample surface which also contribute an electrostatic phase shift. This is a situation that frequently arises when performing cryogenic LTEM. For these cases, the optional amplitude reconstruction branch can be enabled. A second DIP is then trained to output an amplitude map that corresponds to surface contaminants, which is both input directly into the forward model and also scaled to generate the $\phi_e$ induced by the contaminants. This method works because a non-uniform amplitude is required to accurately reproduce the very low image intensity that occurs at the location of surface contaminants, and from this amplitude map we can also approximate $\phi_e$, which is added to $\phi_m$ to give the total phase shift that is input to the forward model. The amplitude DIP thus learns to generate the image contrast from the contaminants while the phase DIP learns to output the phase shift corresponding to the magnetic contrast. We have chosen to focus on the case of surface contaminants that generate both a non-uniform amplitude and a contribution to $\phi_e$, but we believe this method could be adapted and extended to account for sources that generate only a non-uniform $\phi_e$, such as variations in local electrostatic potential in multiferroics, or only a non-uniform amplitude, for example bend contours that generate strong amplitude contrast but do not contribute to the Aharonov-Bohm phase shift.

\subsection*{Demonstration of SIPRAD with simulated data}
\begin{figure}[!htb]
\centering
  \includegraphics[width=\linewidth]{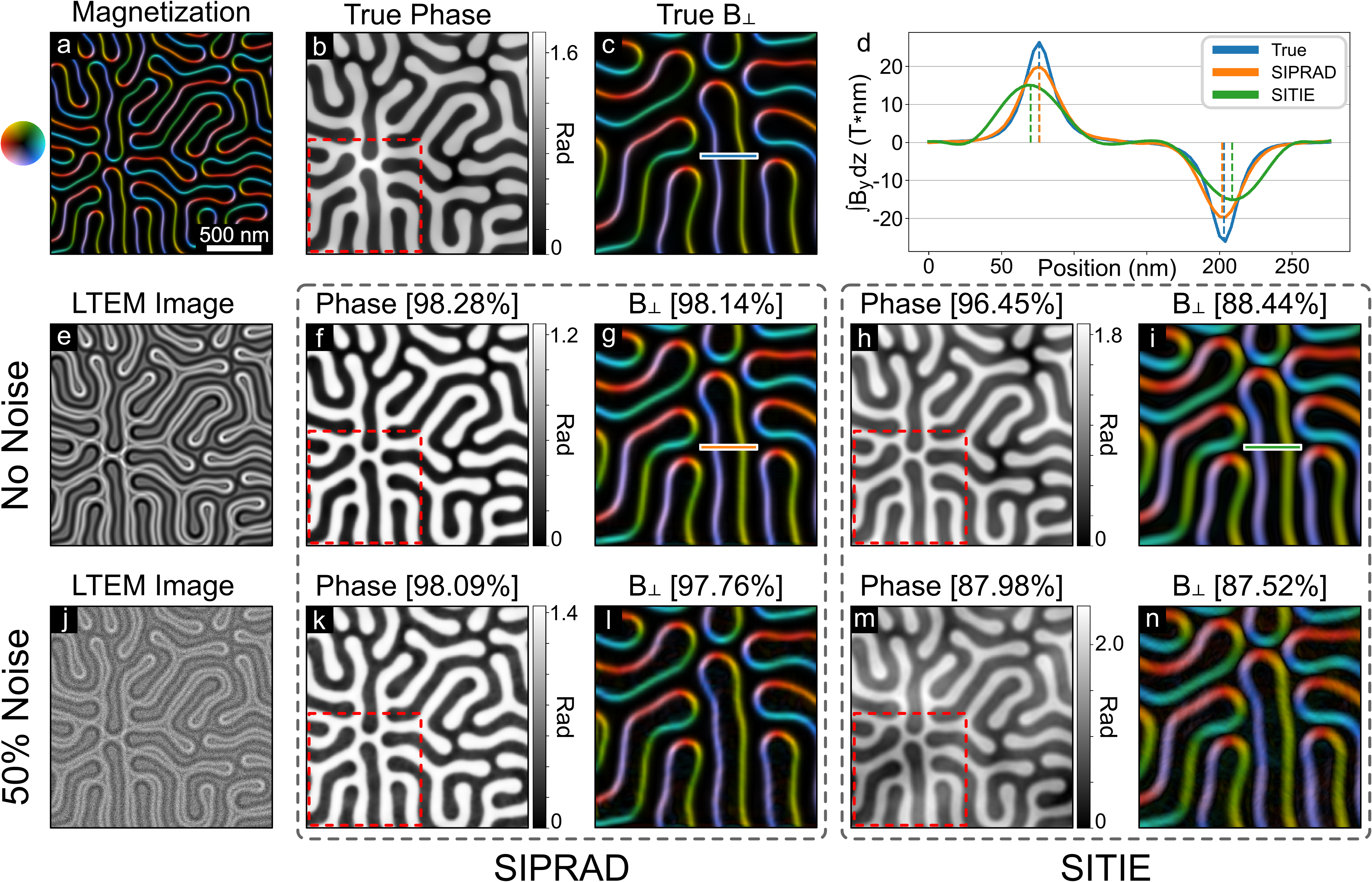}
  \caption{\textbf{Demonstration of SIPRAD and SITIE reconstruction approaches for simulated data.} \textbf{a} In-plane component of the simulated magnetization for stripe domains in CGT. \textbf{b} Ground-truth phase shift corresponding to \textbf{(a)}. \textbf{c} Ground truth in-plane integrated magnetic induction map corresponding to the red boxed region in \textbf{(b)}. \textbf{d} Comparison between line profiles of the $y$ component of the integrated induction across the lines shown in \textbf{(c)}, \textbf{(g)}, and \textbf{(i)}. Dashed lines show domain wall centers. \textbf{e} Simulated LTEM image for $\Delta z = \qty{-500}{\micro \meter}$. \textbf{f} Phase image reconstructed from \textbf{(e)} using SIPRAD. \textbf{g} Integrated in-plane magnetic induction map reconstructed using SIPRAD (of red boxed region).  \textbf{h, i} SITIE-reconstructed phase and in-plane integrated magnetic induction map using \textbf{(e)} as an input image. \textbf{j} LTEM image of same area as shown in \textbf{(e)} but with 50\% Gaussian noise added. \textbf{k-n} Phase reconstructions of the noised image carried out using (k) SIPRAD and (m) SITIE, with corresponding in-plane integrated magnetic induction maps ((l) and (n)). For all images, the number in bracket denotes accuracy compared to the ground truth.}
  \label{fig:demo}
\end{figure}

We first compare the phase retrieval accuracy of SIPRAD and SITIE for stripe domains in a simulated, uniformly thick CGT sample. Figure \ref{fig:demo}a shows the in-plane magnetization component that has been simulated using the Mumax3 micromagnetics package \cite{Vansteenkiste2014}. The calculated electron phase shift (without the uniform electrostatic offset) is shown in Fig.~\ref{fig:demo}b and the integrated in-plane magnetic induction map, $B_\perp$, from the boxed region in b is shown in Fig.~\ref{fig:demo}c. Fig.~\ref{fig:demo}e shows a simulated LTEM image for the phase shift shown in b, with $\Delta z = \qty{-1}{\mm}$, and the corresponding phase reconstructions and integrated induction maps are shown in Fig.~\ref{fig:demo}f-i using the SIPRAD (f, g) and SITIE (h, i) techniques respectively. The accuracy of each reconstruction is quantified as the correlation between the reconstruction and the ground truth and is printed in brackets. For the integrated in-plane magnetic induction, the accuracy of the $x$ and $y$ components are calculated individually and then averaged. 

The SIPRAD approach reconstructs the phase with slightly higher accuracy than the SITIE approach, but both techniques perform well. The integrated induction map, however, is much more accurate for SIPRAD than for SITIE, and this is confirmed in Fig.~\ref{fig:demo}d, which shows a line profile of the $y$ component of the in-plane integrated magnetic induction across two domain walls. The SIPRAD profile matches well to the ground truth both in terms of domain wall location and of width, although the true peak intensity is not reached. The SITIE reconstruction, by contrast, shows a domain wall that is both wider than the true wall and centered in the wrong location. The inaccurate placement of domain walls when using the SITIE method is due to an assumption that ignores the unequal intensities of converging and diverging domain walls, and is a known issue with the technique \cite{Chess2017}. 

We next demonstrate the application of the SIPRAD method to a more experimentally-relevant, noisy input image. Fig.~\ref{fig:demo}j shows the same LTEM image with the addition of \qty{50}{\percent} Gaussian noise, and the corresponding phase reconstructions and integrated magnetic induction maps are shown in Fig.~\ref{fig:demo}k-n. The accuracies of both the phase and the integrated induction remain high for the SIPRAD method, showing only a slight decrease with the addition of noise. The phase accuracy for the SITIE method decreases sharply, and the accuracy of the in-plane integrated magnetic induction is low for both the noise-free and noisy images. The accuracy of the SITIE-reconstructed integrated induction does not significantly decrease because the additional noise manifests in longer-range variations in the phase shift. These are not reflected in the in-plane magnetic induction, which is obtained by taking a gradient of the phase shift. 

For this simple example, SIPRAD outperforms SITIE with respect to the accuracy of both the phase shift and the integrated magnetic induction maps for all cases. The SIPRAD technique accurately determines the location and width of the domain walls, which can be important when trying to use domain information to calculate magnetic material parameters such as the micromagnetic exchange stiffness \cite{lloyd2001, Phatak2015, mccray2022b}.

\subsection*{Increased robustness against noise}
To better compare the accuracy of the SIPRAD and SITIE techniques, we compute the accuracy of both the reconstructed phase and the in-plane integrated magnetic induction for a wide range of defocus and noise values. There are several noise sources that contribute to LTEM images, including Poisson, Gaussian, and salt \& pepper noise. We have found that SIPRAD performs similarly well when given an input with equal amounts of either pure Gaussian noise or a mix of noise types, while SITIE performs slightly better for pure Gaussian noise (see Supplementary Section S1). Gaussian noise was therefore used in the following accuracy measurements as it is easier to quantify. The level of Gaussian noise refers to the standard deviation of the noise distribution relative to the mean image intensity. 

\begin{figure}[htb]
\centering
  \includegraphics[width=0.7 \linewidth]{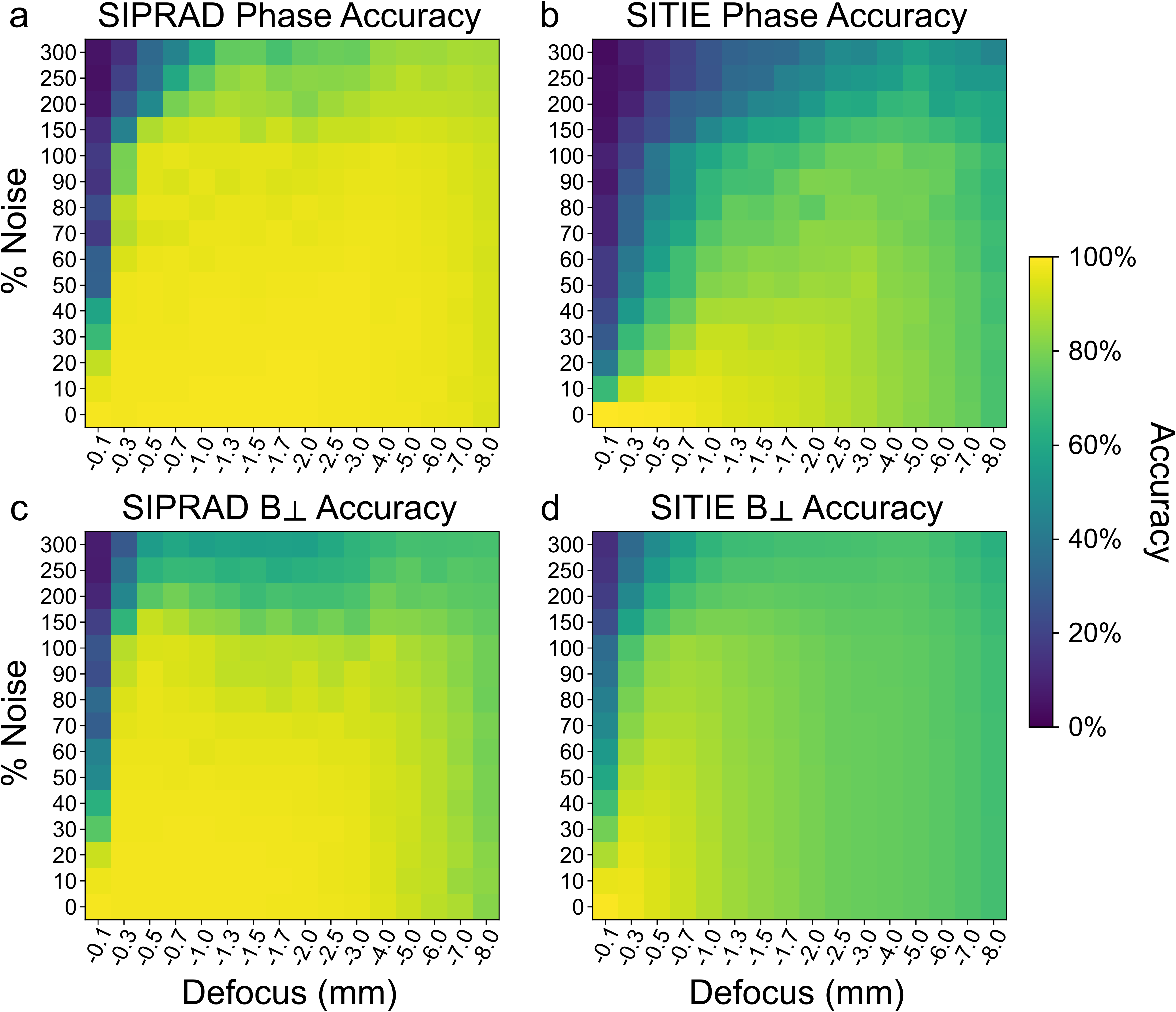}
  \caption{\textbf{Comparison of accuracy between SIPRAD and SITIE. a, b} Phase accuracy for the SIPRAD and SITIE approaches, plotted on a color scale as a function of \% noise and defocus of the input image. \textbf{c, d} Accuracy of the in-plane integrated magnetic induction displayed in the same way as for \textbf{(a)} and \textbf{(b)}.}
  \label{fig:acc}
\end{figure}

Figure \ref{fig:acc} shows the accuracy of the SIPRAD and SITIE techniques applied to an input image simulated from the true phase in Fig.~\ref{fig:demo}, for defocus values ranging from \qty{-0.1}{\mm} to \qty{-8}{\mm}, and noise values ranging from \qty{0}{\percent} to \qty{300}{\percent}. The SIPRAD phase reconstruction is very accurate across a large portion of the parameter space (yellow region), while at high defocus values the accuracy of the in-plane integrated magnetic induction reduces somewhat. This is caused by the extreme blurring that occurs in images with very high defocus. The SITIE reconstructions, by contrast, are only accurate within a narrow region of low defocus and low noise. 

For both reconstruction methods, noise is more impactful at low defocus values. This is because the intensity of the magnetic contrast increases with increased defocus. At $\Delta z = \qty{-100}{\micro \meter}$, even small amounts of noise can completely obscure the magnetic contrast in the image, which is why real-world LTEM imaging is normally performed at relatively high defocus values, outside the small defocus limit within which the SITIE approach is accurate. The SIPRAD technique, by contrast, shows an increase in accuracy up to, and beyond, a defocus of \qty{-1}{\mm}, meaning that it is most accurate for real-world imaging conditions. Likewise, the SIPRAD technique maintains a high accuracy in both the reconstructed phase images and the in-plane integrated magnetic induction maps up to noise values approaching 100\%. Both of these trends are also reflected when measuring accuracy using the structural similarity index measure (SSIM), as shown in Supplementary Section S2. 

The SIPRAD method is accurate when reconstructing the phase shift from images taken at high defocus, which means that it can be applied to samples with weak magnetic moments or which would otherwise induce only a small magnetic phase shift. Due to the high noise tolerance, shorter acquisition times can be used which makes SIPRAD viable for imaging highly beam-sensitive samples and for imaging with increased time resolution. 

\subsection*{Phase reconstruction with amplitude variation}
The most important development enabled by our SIPRAD approach is not the increased accuracy or robustness to noise, but the ability to perform magnetic phase reconstructions on samples across which the amplitude and electrostatic phase shift of the exit wave are varying. All of the previously-introduced phase retrieval techniques, including the TIE approach and off-axis holography, reconstruct the total electron phase shift $\phi$ of the sample. If the electrostatic phase shift, $\phi_e$, is uniform across the sample, then the gradients of $\phi$ and $\phi_m$ are equal and $B_\perp$ can be obtained. However, if the mean inner potential or the sample thickness is not uniform, then the amplitude of the electron wave and $\phi_e$ are not constant and a single phase reconstruction cannot be used to calculate $B_\perp$. There are methods for separating $\phi_e$ and $\phi_m$, such as performing a second phase reconstruction with the sample inverted in the microscope \cite{Humphrey2014}, but this is not often applicable during in situ experiments as this requires the sample to be fully removed from the microscope \cite{pollard2012}.

The distinction between $\phi$ and $\phi_m$ is especially important when considering SITIE reconstructions. A major assumption for the SITIE technique is that all contrast in the defocused image arises from magnetic contributions in the sample. This requires samples that do not display diffraction or amplitude contrast in the images, and that are perfectly clean and of uniform thickness. These conditions are frequently not met. It is especially difficult to ensure these conditions when imaging single-crystal samples that generate strong diffraction contrast, or when performing cryogenic in situ experiments that can easily lead to sample surface contamination. Contamination that collects on the sample is particularly problematic, as it can create variations in both the amplitude and the electrostatic phase shift. While diffraction contrast can frequently be avoided by adjusting sample tilt and imaging conditions, contrast from contamination cannot be removed once the sample is in the microscope. Especially troubling is that surface contamination can lead to phase reconstructions and to integrated magnetic induction maps with features that appear similar to viable magnetic features, but which are actually non-magnetic in origin.

Performing the amplitude reconstruction poses an additional computational problem, namely that an amplitude map can be created that generates contrast which matches magnetic contrast. The reconstruction algorithm is therefore less stable when performing amplitude and $\phi_e$ reconstructions and will often diverge or generate non-physical amplitude maps. We remedy this by adding the additional constraint that the amplitude should be dual-valued, i.e. uniform except in finite regions that correspond to the surface contamination (which is visible in the bright-field, in-focus image), and this sufficiently constrains the model such that reconstructions can be performed in many cases. The general workflow of the algorithm does not change when performing amplitude reconstructions, but the additional branch shown in Fig.~\ref{fig:outline} is included. The input into the amplitude DIP is a thresholded version of the input image. This is created with a manually-chosen threshold value that approximately isolates the contaminants. The output of the amplitude DIP is then used directly as the amplitude in the forward model. This output amplitude is scaled according to user-input material parameters to convert it to $\phi_e$ (for cases where both amplitude and $\phi_e$ are locally modified), which is added to the value of $\phi_m$ from the phase DIP to give the total phase shift input to the forward model. When backpropagating, both DIPs are optimized with every iteration. Other training methods, including alternating between optimizing the phase and amplitude DIPs, made the algorithm less accurate and stable. 

\begin{figure}[!ht]
\centering
  \includegraphics[width=\linewidth]{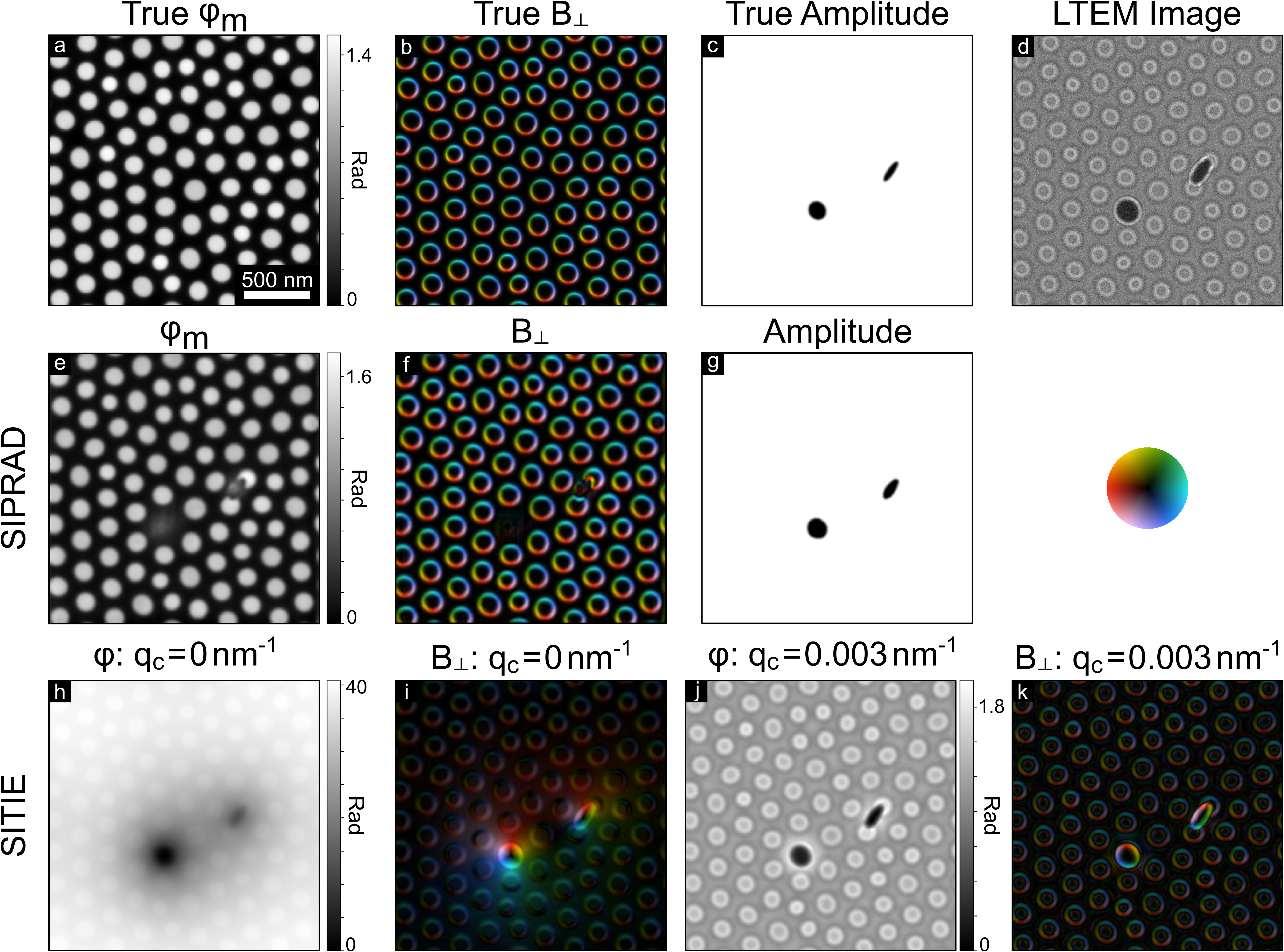}
  \caption{\textbf{Application of phase reconstruction to samples with surface contamination. a} True magnetic phase shift and \textbf{b} integrated magnetic induction map showing simulated bubble domains in CGT. \textbf{c} True amplitude map including two pieces of surface contamination that impart both amplitude variation and electrostatic phase shift. \textbf{d} Resulting LTEM image with 30\% noise added. \textbf{e} SIPRAD-reconstructed magnetic phase shift and \textbf{f} corresponding integrated magnetic induction map. \textbf{g} SIPRAD-reconstructed amplitude map. SITIE-reconstructed maps of \textbf{h} the total phase shift, and \textbf{i} the in-plane magnetic induction for the same input image. \textbf{j, k} SITIE-reconstructed maps of the total phase shift and of the magnetic induction with a manually-optimized Tikhonov filter applied.}
  \label{fig:amp}
\end{figure}

Figure \ref{fig:amp} shows an application of amplitude learning to a simulated sample of CGT containing bubble domains and with added surface contamination. Fig.~\ref{fig:amp}a and b show the true magnetic phase shift, $\phi_m$, and the corresponding in-plane integrated magnetic induction map. Two regions of surface contamination are then added that contribute to a non-uniform electrostatic phase shift, $\phi_e$, and the corresponding amplitude map is shown in Fig.~\ref{fig:amp}c. An LTEM image, simulated with $\Delta z = \qty{-500}{\micro \meter}$ and 30\% added noise, is shown in Fig.~\ref{fig:amp}d; the surface contamination is clearly visible. Fig.~\ref{fig:amp}e-g show the SIPRAD-reconstructed map of $\phi_m$, the in-plane integrated magnetic induction map, and the amplitude map, respectively. The amplitude matches the ground truth well, and the phase map is accurate except in regions obscured by the surface particles. There are some artifacts in the magnetic induction map near the edge of the surface particles, but there is otherwise a very good match to the ground truth. 

By contrast, the SITIE approach does not produce an accurate reconstruction of either $\phi_m$ or $B_\perp$ from the LTEM image shown in d. Figs. \ref{fig:amp}h and i show the SITIE-reconstructed phase shift (h) and integrated magnetic induction map (i) from the same input image (d). The scale of the phase shift is very inaccurate due to the additional contrast from the amplitude and $\phi_e$ introduced by the surface particles. The particles also dominate the integrated magnetic induction maps, and the actual magnetic domains are not visible. One can improve the quality of the SITIE reconstruction by applying a Tikhonov filter, as shown in Fig.~\ref{fig:amp}j for a manually-optimized Tikhonov frequency of $q_c = \qty{0.003}{\per \nano \meter}$. The Tikhonov filter makes the magnetic components of the phase shift more visible, but the image is still both quantitatively and qualitatively inaccurate. The magnetic information in the in-plane integrated induction map shown in Fig.~\ref{fig:amp}k is also improved, but the false contrast from the surface particles still dominates, and in this case could be easily mistaken for a magnetic bubble with opposite chirality to that of the other magnetic bubbles.

\subsection*{Application to experimental data}
\begin{figure}[htb]
\centering
  \includegraphics[width=\linewidth]{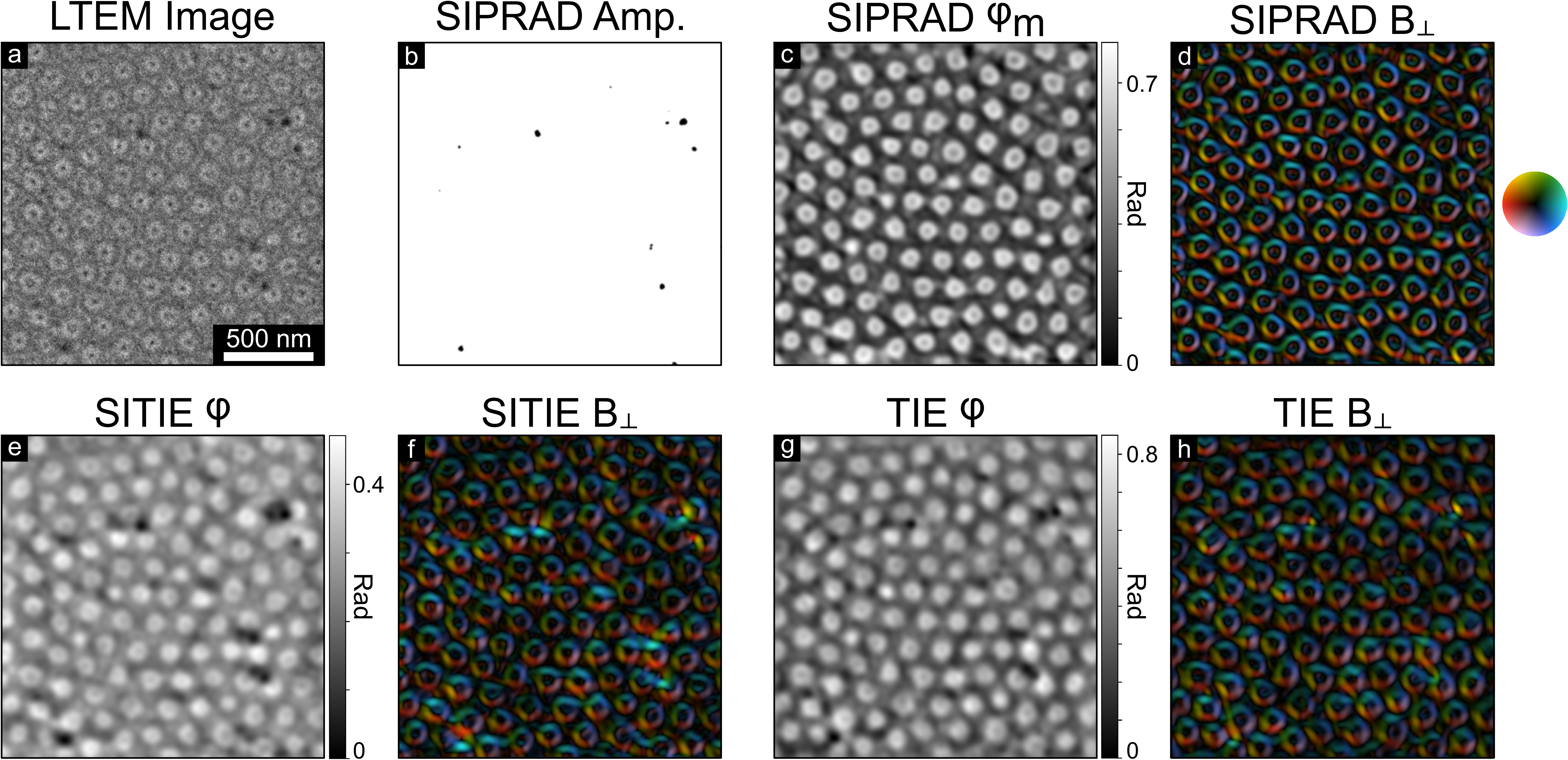}
  \caption{\textbf{Application of SIPRAD to experimental data. a} Experimental LTEM image of bubble domains in CGT recorded with $\Delta z = \qty{-700}{\micro \meter}$ and at \qty{23}{\kelvin}. \textbf{b} Amplitude map obtained from thresholding the input image to separate out the surface particles visible in \textbf{a}. \textbf{c, d} SIPRAD-reconstructed maps of \textbf{c} the magnetic phase shift, and \textbf{d} the in-plane integrated magnetic induction. \textbf{e, f} SITIE-reconstructed maps of the total phase shift and induction map from \textbf{(a)}. \textbf{g, h} TIE-reconstructed maps of the total phase shift and the integrated magnetic induction using a TFS with $\Delta z = \pm \qty{700}{\micro \meter}$.}
  \label{fig:exp}
\end{figure}

We now apply the SIPRAD reconstruction technique to an experimental LTEM image of an exfoliated CGT flake, shown in Figure \ref{fig:exp}a. The sample was field cooled through the Curie temperature to \qty{23}{\kelvin} in a \qty{500}{\Oe} field to stabilize magnetic bubble domains, and was then imaged using a defocus of $\Delta z = \qty{-700}{\micro \meter}$. The experimental image shows clear bubble domains along with additional contrast due to surface contamination. The image is very noisy, and the SIPRAD algorithm will often diverge when trying simultaneously to reconstruct both the amplitude and phase for this image. The contrast from the contaminants can still be taken into account, however, by using a fixed, non-uniform amplitude map and corresponding electrostatic phase shift, $\phi_e$. We first approximate the amplitude map by manually thresholding the input image to create the binary image shown in Fig.~\ref{fig:exp}b, which is used to obtain the amplitude function and scaled to create $\phi_e$. These components of the wave function contribute to the forward model but remain fixed during the reconstruction process that optimizes the magnetic phase shift, $\phi_m$. The training process is thus stable while still taking into account local contributions to the amplitude and to $\phi_e$, which allows the reconstruction of $\phi_m$ isolated from the total phase shift, $\phi$. Fig.~\ref{fig:exp}c and d show the SIPRAD-reconstructed maps of $\phi_m$ and of the in-plane integrated magnetic induction, respectively. The reconstructed map of $\phi_m$ matches well to the simulated phase map of bubble domains shown in Fig.~\ref{fig:amp}a. The reconstruction shows inaccuracies around the areas of surface contamination, where the magnetic information is lost. Errors are also observed near the edge of the image due to periodic boundary conditions enforced by the forward model. Performing the same reconstruction with a uniform amplitude map leads to a significantly worse reconstruction with visible artifacts from the surface particles (Supplementary Section S3). 

Fig.~\ref{fig:exp}e and f show maps of the SITIE-reconstructed total phase shift and of the in-plane integrated magnetic induction from the same experimental image (shown in a) with an optimized value of the Tikhonov filter frequency, $q_c$. The phase map is dominated by the contribution from the surface particles, which lead to artifacts in the induction map, such as varying intensities for different bubbles and an apparent bubble with opposite chirality to the others. We also performed a TIE reconstruction of the same region using a TFS of three images taken at defocus values of $\Delta z = \pm \qty{700}{\micro \meter}$ and $\Delta z = \qty{0}{\micro \meter}$. Note that the underfocus image from the TFS is what was used for the SIPRAD and SITIE reconstructions. The full TFS is shown in Supplementary Section S4 and surface contaminants are clearly visible in the in-focus image. An optimal value of $q_c$ was determined manually as for the SITIE reconstruction. The TIE-reconstructed phase shift map is significantly improved compared to the map obtained for the SITIE reconstruction, primarily because the amplitude contrast from the surface particles is present in all images in the TFS and is thus accounted for, but the contribution from $\phi_e$ is still evident. 

Even though the SIPRAD approach uses only one image, rather than a TFS of three images, the SIPRAD-reconstructed maps of the phase shift and integrated magnetic induction both appear qualitatively to better represent the true domain structure than those produced by the TIE reconstruction. This is further supported by comparing the domain wall profiles in the reconstructed magnetic induction maps to the true profile of a simulated bubble, which is shown in Supplementary Section S5. The SIPRAD reconstruction with a fixed, dual-valued amplitude map therefore appears to enable the magnetic phase shift to be quantitatively and qualitatively determined from the experimental data shown in Fig.~\ref{fig:exp}. 

\begin{figure}[htb]
\centering
  \includegraphics[width=\linewidth]{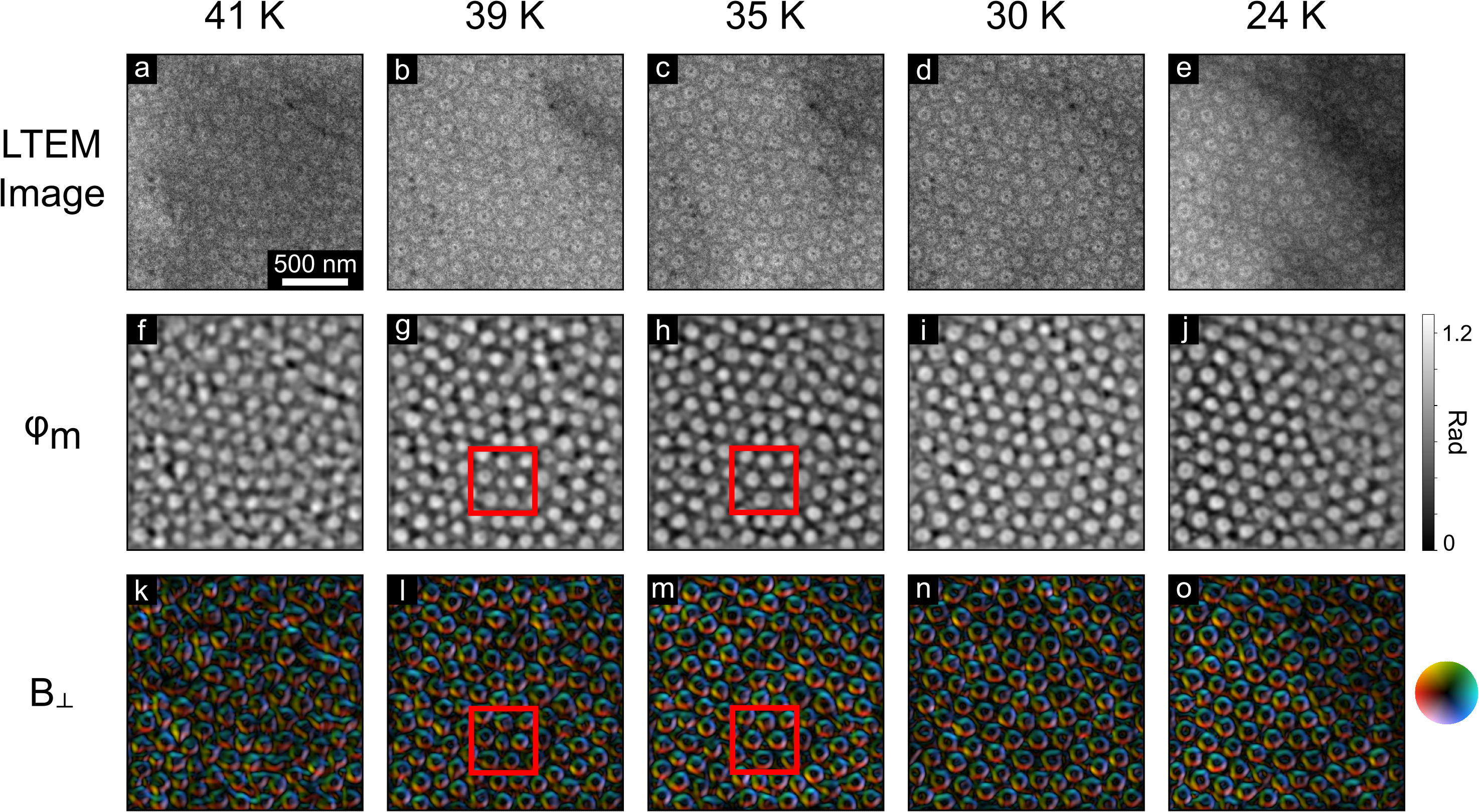}
  \caption{\textbf{Application of SIPRAD to an in situ field-cooling experiment. a-e} Experimental LTEM images of CGT during field cooling in a \qty{500}{\Oe} field, recorded with $\Delta z = \qty{-700}{\micro \meter}$. Temperatures are shwn above each image. \textbf{f-j} Magnetic phase reconstructions performed with the SIPRAD method of the images shown in \textbf{a-e}. Red boxes highlight an area where a bubble has disappeared between \qty{39}{\kelvin} and \qty{35}{\kelvin}. \textbf{k-o} Integrated magnetic induction maps generated from the magnetic phase reconstructions shown in \textbf{f-j}.}
  \label{fig:insitu}
\end{figure}

We next show the application of SIPRAD to experimental in situ LTEM cooling data. This is an example of a situation in which acquiring a TFS of images is not feasible. Figure \ref{fig:insitu}a-e show a series of LTEM images of a CGT flake recorded during field-cooling of the sample from above the Curie temperature to \qty{24}{\kelvin}. As the temperature is reduced, the sample expands due to magnetostrictive effects; this leads to the formation of bend contours that can obscure the magnetic contrast, as can be seen in Fig.~\ref{fig:insitu}e \cite{mccray2023}. 

Fig.~\ref{fig:insitu}f-j shows the SIPRAD-reconstructed phase shift maps. The corresponding integrated induction maps are shown in Fig.~\ref{fig:insitu}k-o. The phase and induction maps allow easy visualization of how the magnetic bubbles rearrange when cooling, and show that some bubbles are driven out of the sample, as highlighted by the red boxes in Fig.~\ref{fig:insitu}g and h. Between each temperature step, the magnetic bubble lattice shifts and distorts due to both the disappearance of bubbles and Brownian motion \cite{yao2020, suzuki2021}. The SIPRAD method enables quantitative analysis of the bubble-lattice time evolution, because phase reconstruction allows accurate tracking of bubbles between sequential images and gives further insight into local changes of the magnetic domain structure.

\section*{Discussion}
In this work, we have demonstrated that we can reconstruction the magnetic phase shift, and thus the integrated magnetic induction, from a single input LTEM image, using a DIP-enabled automatic differentiation method. The SIPRAD approach is shown to be significantly more accurate than existing techniques over a wide range of input conditions, including moderate to large defocus values and noisy inputs that best represent experimental data. The use of a large defocus is often required when performing experimental LTEM, as the increased phase sensitivity at large defocus allows magnetic contrast to be more easily distinguished from noise and other contrast contributions. Other phase reconstruction methods, such as those based on solving the TIE, are only accurate in the small defocus limit and lose both spatial and phase resolution when experimentally-relevant defocus values are used. The SIPRAD method, by contrast, is based on a forward model that retains high accuracy of the reconstructed phase to defocus values as large as \qty{-6}{\milli \meter}, and accuracy of the in-plane integrated magnetic induction to defocus values as large as \qty{-3}{\milli \meter}. The robustness of the approach against high noise levels is primarily due to the implementation of a DIP. DIPs have been shown to be an effective way of regularizing noisy input data without relying on manually-chosen parameters, and this work further reinforces these findings. The downside of the using a DIP in AD-based phase reconstruction is that computationally it is much slower than inference by a conventional machine-learning model, though with the benefit of not needing to train such a model in the first place. Reconstructing each image requires optimizing a new CNN, but we find that the $\sim 5000$ iterations required to converge the model for a 512 x 512 image takes only 35 seconds on an NVIDIA A100 GPU. We are able to achieve this reconstruction time by inputting an estimated phase shift calculated with the SITIE method, which reduces the total number of iterations required. By demonstrating the high accuracy at both large defocus and low signal-to-noise levels, we have shown that SIPRAD is an excellent technique for reconstructing LTEM images in these experimental conditions, such as would occur when imaging weakly magnetic or beam-sensitive samples. 

Although the increased accuracy and robustness towards noise is promising, the most important aspect of the SIPRAD approach is that it allows $\phi_m$ to be reconstructed even in the presence of image contrast arising from variations in amplitude and $\phi_e$. Real-world samples inevitably contain heterogeneities that can make magnetic phase reconstruction challenging. This is especially true when imaging spin textures across a large field of view. We have demonstrated that the SIPRAD approach enables simultaneous reconstruction of both the amplitude and $\phi_m$, but we find that to optimize both components from a single image can be difficult and will often lead the algorithm to diverge at high noise levels. Even in these cases, however, the magnetic phase can still be effectively isolated and reconstructed after creating a thresholded binary amplitude map (or otherwise determining the amplitude distribution), as demonstrated in Fig.~\ref{fig:exp}. This allows for reconstructing phase maps and in-plane integrated magnetic induction maps from images for which the TIE method and a single TFS do not work. This also provides an example of the adaptability of the SIPRAD method. Known information about the sample, such as topographical data for patterned nanostructures, could be additionally included into the forward model and used to constrain the reconstruction and improve the isolation of $\phi_m$. 

The fact that SIPRAD can accommodate large defocus values, high noise levels, and heterogeneous samples or those with surface particulates demonstrates that it is most useful when given noisy images recorded in conditions that do not allow for the sample to be inverted or even a TFS to be recorded. This makes the technique especially appealing for application to in situ experiments. When studying the time evolution of magnetic domains it is frequently impossible to collect a TFS at each point of interest--normally a movie of a series of defocused images will be the only data available. SIPRAD is a viable technique able to perform both quantitatively- and qualitatively-accurate reconstructions of $\phi_m$ and of the integrated magnetic induction from single movie frames. In addition, in these scenarios it is often paramount to study the evolution of one particular region over time, and SIPRAD is able to isolate $\phi_m$ even for changes in the imaging conditions, or additional surface features, that develop during the experiment. We therefore hope that SIPRAD can be broadly applied to future LTEM in situ experiments and studies of difficult-to-observe magnetic spin textures.

\section*{Methods}

\subsection*{Single image phase reconstruction via automatic differentiation}
SIPRAD is implemented in PyTorch using the Adam optimizer. The DIPs are convolutional autoencoders as depicted in Supplementary Section S6. The input for the phase DIP is a SITIE phase reconstruction of the input image, and the input for the amplitude DIP is a thresholded version of the input image with the threshold value manually set to isolate features such as surface contamination. Similar to other implementations, we augment our input image by adding small amounts of noise in each iteration \cite{ulyanov2018}. Each DIP is pre-trained for 200 iterations to produce its input image before the full algorithm is run with the forward model. We use a convolutional autoencoder deep neural network for our DIPs, and the architecture is outlined graphically in Supplementary Section S6. The final reconstructed $\phi_m$ is smoothed by a Gaussian filter with $\sigma = 2$ pixels. 

When quantifying the accuracy for Fig.~\ref{fig:acc}, 10 randomly-noised images (with the same noise level) were created for each defocus value and used for the phase reconstructions. The average accuracy of the 10 reconstructions of the phase maps and the integrated magnetic induction maps is displayed for both SIPRAD and SITIE.  

When training with amplitude reconstruction, the dual-valued constraint is implemented every 100 iterations of the reconstruction as well as after the final iteration. The high and low values are determined from the mode and minimum of all pixels, and the threshold value is chosen as 3/4 of the mode.  Early stopping is sometimes necessary, especially when performing amplitude reconstruction, to prevent over-fitting. The amplitude map that we are reconstructing is based on the thickness and mean inner potential of the material, as described in the forward model section. 

\subsection*{Experimental dataset}
Experimental cryo-LTEM imaging of CGT was performed on a JEOL JEM-2100F TEM instrument using a Gatan double-tilt liquid helium holder. TEM samples were created by dry-exfoliating flakes from a bulk crystal, which were then placed on a silicon nitride TEM membrane. The thickness of the flake was measured with atomic force microscopy to be 150 nm. The sample was field-cooled in the microscope in a \qty{500}{\Oe} out-of-plane magnetic field to \qty{23}{\kelvin} in order to nucleate and stabilize magnetic bubbles.

\subsection*{Phase retrieval using the transport of intensity equation}
The transport of intensity equation can be written as 
\begin{equation}
    \nabla \cdot \left( I_0 \nabla \phi \right) = - \frac{2 \pi }{\lambda } \frac{\partial I}{\partial z},
\end{equation}
where $I_0$ is the in-focus image, $\lambda$ the electron wavelength, $\partial I / \partial z$ the through-focal image intensity derivative, with $z$ being the electron propagation direction. Under the single-image TIE approximation, the equation becomes 
\begin{equation}
    \nabla^2 \phi = - \frac{2 \pi }{\lambda \Delta z} \left( \frac{1 - I_{\Delta z}}{I_0} \right),
\end{equation}
where $\Delta z$ is the defocus and the in-focus image $I_0$ is approximated as a uniform image with the mean intensity of the defocused image $I_{\Delta z}$ \cite{Chess2017}. 

The TIE and SITIE methods were implemented for both the simulated and experimental data using the open-source PyLorentz software \cite{Mccray2021, PyLorentzGitHub}. 

\subsection*{LTEM image formation model}
Here we show how the LTEM images are calculated. We begin with the Aharonov-Bohm equation \cite{Aharonov1961}, which describes how electrons that pass through the sample are subject to a phase shift induced by the electrostatic potential ($V$) and magnetic vector potential ($\mathbf{A}$), 

\begin{align}
    \phi &= \phi_e + \phi_m,\nonumber\\
    &= \frac{\pi}{\lambda E} \int V d\mathbf{l} - \frac{\pi}{\varphi_0} \int \mathbf{A} \cdot d\mathbf{l},
\end{align}
where $\phi$ is the total phase shift, $\phi_e$ and $\phi_m$ the electrostatic and magnetic components of the phase shift respectively, $\lambda$ is the electron wavelength, $E$ the relativistic electron energy, $\mathbf{l}$ the electron propagation direction, and $\varphi_0=h/2e$ the magnetic flux quantum. The in-plane component of the integrated magnetic induction can be obtained from $\phi_m$, 
\begin{equation}
    \left( B_x, B_y \right) = \frac{\varphi_0}{\pi} \left( - \frac{\partial \phi_m }{\partial y} , \frac{\partial \phi_m }{\partial x} \right). 
\end{equation}

The total phase shift is used to write the electron wave function at the exit surface of the sample as 
\begin{equation}
    \psi(\rp) = a(\rp) \mathrm{e}^{\mathrm{i} \phi(\rp)},
\end{equation}
where $a(\rp)$ is an amplitude function and $\phi$ is the total electron phase shift. We do not include detailed electron scattering when calculating the amplitude function, but only the absorption of the electrons, which is given by 
\begin{equation}
    a(\rp) = \mathrm{e}^{-t_\perp/\xi_0},
\end{equation}
where $t_\perp$ is the sample thickness shape function, and $\xi_0$ is the absorption coefficient for the sample material.

In order to calculate the resulting image intensity, we propagate this wave function to the exit plane by convolving it with the transfer function of the microscope in the back focal plane and determining the intensity of the wave function,
\begin{equation} \label{eq.im_intensity}
    I(\rp) = \abs{\psi(\rp) * \mathcal{T}(\rp)}^2,
\end{equation}
where $*$ is a convolution operation, and $\mathcal{T}(\rp)$ is the microscope transfer function. This operation is written here in real space but is computed in Fourier space. The LTEM transfer function is composed of three parts \cite{DeGraef2000}:
\begin{equation}
    \mathcal{T}(\kp) = A(\kp) \mathrm{e}^{-\mathrm{i} \chi (\kp)} \mathrm{e}^{-g(\kp)},
\end{equation}
where $A(\kp)$ is the objective aperture function, $\mathrm{e}^{-\mathrm{i} \chi (\kp)}$ the phase transfer function, $\mathrm{e}^{-g(\kp)}$ the damping envelope, and $\kp$ the reciprocal space wave vector perpendicular to the beam direction. The aperture is a binary function (1 inside and 0 outside) centered in reciprocal space for the Fresnel imaging mode. We can define the phase transfer function as 
\begin{equation}
    \chi(\kp) = \pi \lambda \left[ \Delta z + C_a \cos\left(2\phi_a\right) \right] \abs{k}^2 + \frac{\pi}{2} C_s \lambda^3 \abs{k}^4,
\end{equation}
where $\Delta z$ is the defocus, $C_a$ and $\phi_a$ are the magnitude and orientation of the two-fold astigmatism, and $C_s$ is the spherical aberration coefficient. 

The damping envelope can be written as 
\begin{eqnarray}
    g(\kp) =  \frac{\pi^2 \theta_c^2}{\lambda^2 \, u}\left( C_s \lambda^3 \abs{k}^3 - \Delta z\lambda \abs{k}\right)^2  + \frac{\left( \pi \lambda \Delta \right)^2}{2 u}\abs{k}^4, 
\end{eqnarray}
where $u=1+2\left(\pi \theta_c\Delta\right)^2 \abs{k}^2$, $\theta_c$ is the beam divergence angle, and $\Delta$ is the defocus spread \cite{Graef}. LTEM image simulations are performed using the PyLorentz software \cite{Mccray2021, PyLorentzGitHub}.

\subsection*{Simulated datasets}
Micromagnetic simulations of CGT were performed with Mumax3 \cite{Vansteenkiste2014}. The following parameters were used: cell size $\left( \qty{4}{\nano \meter} ~ \times ~ \qty{4}{\nano \meter} ~ \times ~ \qty{10}{\nano \meter} \right)$, grid size $\left( \mathrm{512 ~ \times ~ 512 ~ \times ~ 10} \right)$, $M_s = \qty{2E5}{\ampere \per \meter}$, $A_{ex} = \qty{1.2 E-12}{\joule \per \meter}$, $K_u = \qty{2E4}{\joule \per \meter \cubed}$, $B_{ext} = \qty{0.08}{\tesla}$ along the $z$ direction. For simulations of magnetic bubbles, a random starting magnetization of Bloch type domains and bubbles was placed and then relaxed. The PyLorentz software was used to calculating the total electron phase shift of the simulated samples \cite{Mccray2021, PyLorentzGitHub}. Gaussian noise was added to the images after they are created. The level of Gaussian noise refers to the standard deviation of the distribution divided by the mean intensity of the image.

\section*{Data Availability}
Data and code will be made available in a public Github repository. 

\bibliography{references}

\section*{Acknowledgments}
This work was supported by the U.S. Department of Energy, Office of Science, Office of Basic Energy Sciences, Materials Sciences and Engineering Division. This work was performed, in part, at the Center for Nanoscale Materials and the Advanced Photon Source, both U.S. Department of Energy Office of Science User Facilities, and supported by the U.S. Department of Energy, Office of Science, under Contract No. DE-AC02-06CH11357. M.J.C also acknowledges support from Argonne LDRD 2021-0090 – AutoPtycho: Autonomous, Sparse-sampled Ptychographic Imaging. We gratefully acknowledge the computing resources provided on Swing, a high-performance computing cluster operated by the Laboratory Computing Resource Center at Argonne National Laboratory. We would like to acknowledge Yue Li for her help in acquiring the experimental data.

\section*{Author contributions statement}
C.P. and M.J.C. conceived the project. A.M implemented the SIPRAD algorithm using the PyTorch package, generated the simulated data, analyzed the results, and wrote the manuscript. A.M. and C.P. generated the experimental data. C.P., M.J.C., and A.P. supervised the project. All authors discussed the results and contributed to the paper.

\section*{Competing Interests}
The authors declare no competing interests.

\section*{Additional information}
Supplementary information is available. 

\end{document}


\maketitle

\section{Mixed noise images}

\begin{figure}[htb]
\centering
  \includegraphics[width=\linewidth]{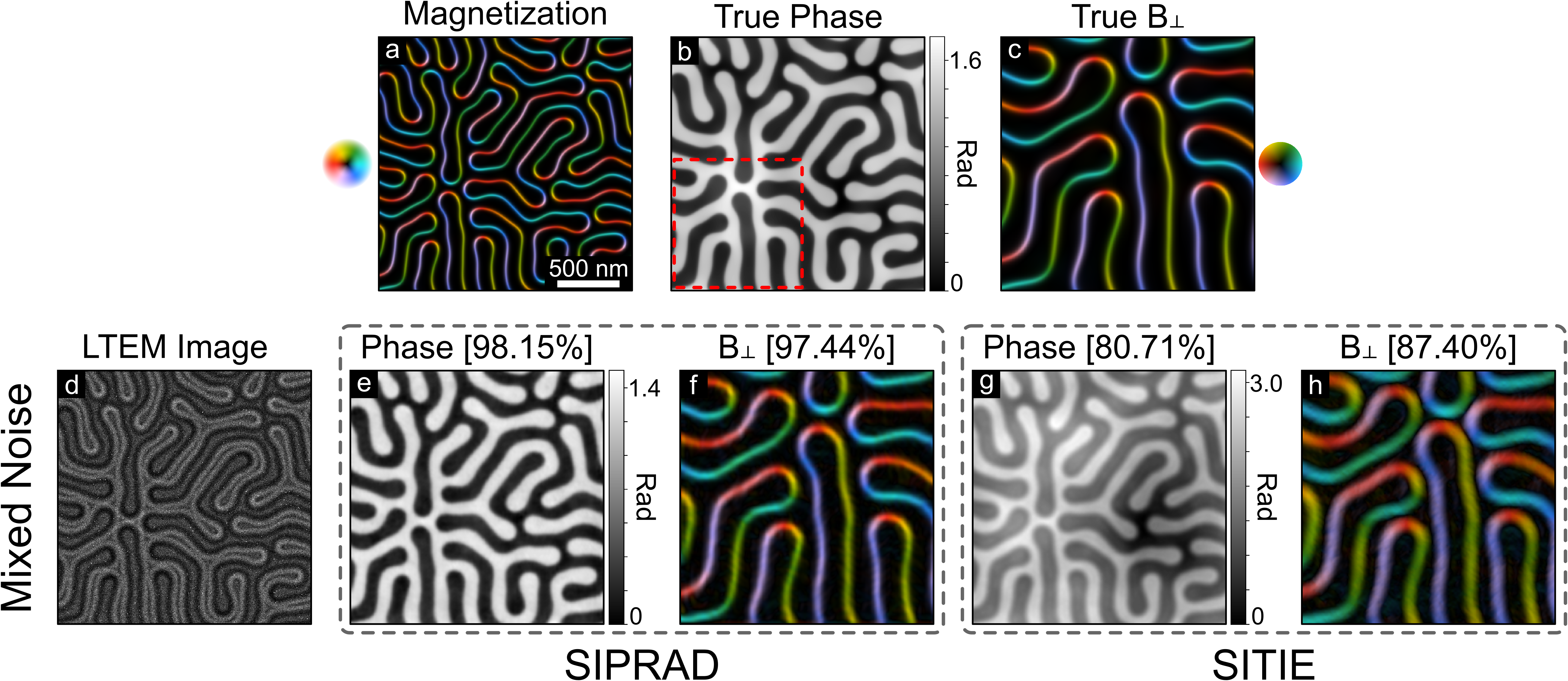}
  \caption{\textbf{Accuracy of SIPRAD and TIE with mixed noise images. a-c} True magnetization, phase shift, and in-plane induction map of a simulated CGT sample, identical to main text Figure 2. All induction maps correspond to the red-boxed region shown in \textbf{(b)}. \textbf{d} LTEM image simulated from \textbf{(b)} at $\Delta z = \qty{-500}{\micro \meter}$, with a mixture of Gaussian, Poisson, and salt-and-pepper noise added. \textbf{e} SIPRAD reconstructed phase from \textbf{(d)}; for all images, number in bracket denotes accuracy compared with ground truth. \textbf{g, h} SITIE reconstructed phase and integrated induction map using \textbf{(e)} as an input image. 
  }
  \label{figSup:mix}
\end{figure}

\clearpage
\section{SIPRAD vs SITIE SSIM}
Figure 3 in the main text plots correlational accuracy of phase and induction maps reconstructed with the SIPRAD and SITIE methods. Here we show that the results given there still hold when using a different measure of accuracy, namely the structural similarity index measure (SSIM). 

\begin{figure}[htb]
\centering
  \includegraphics[width=0.7 \linewidth]{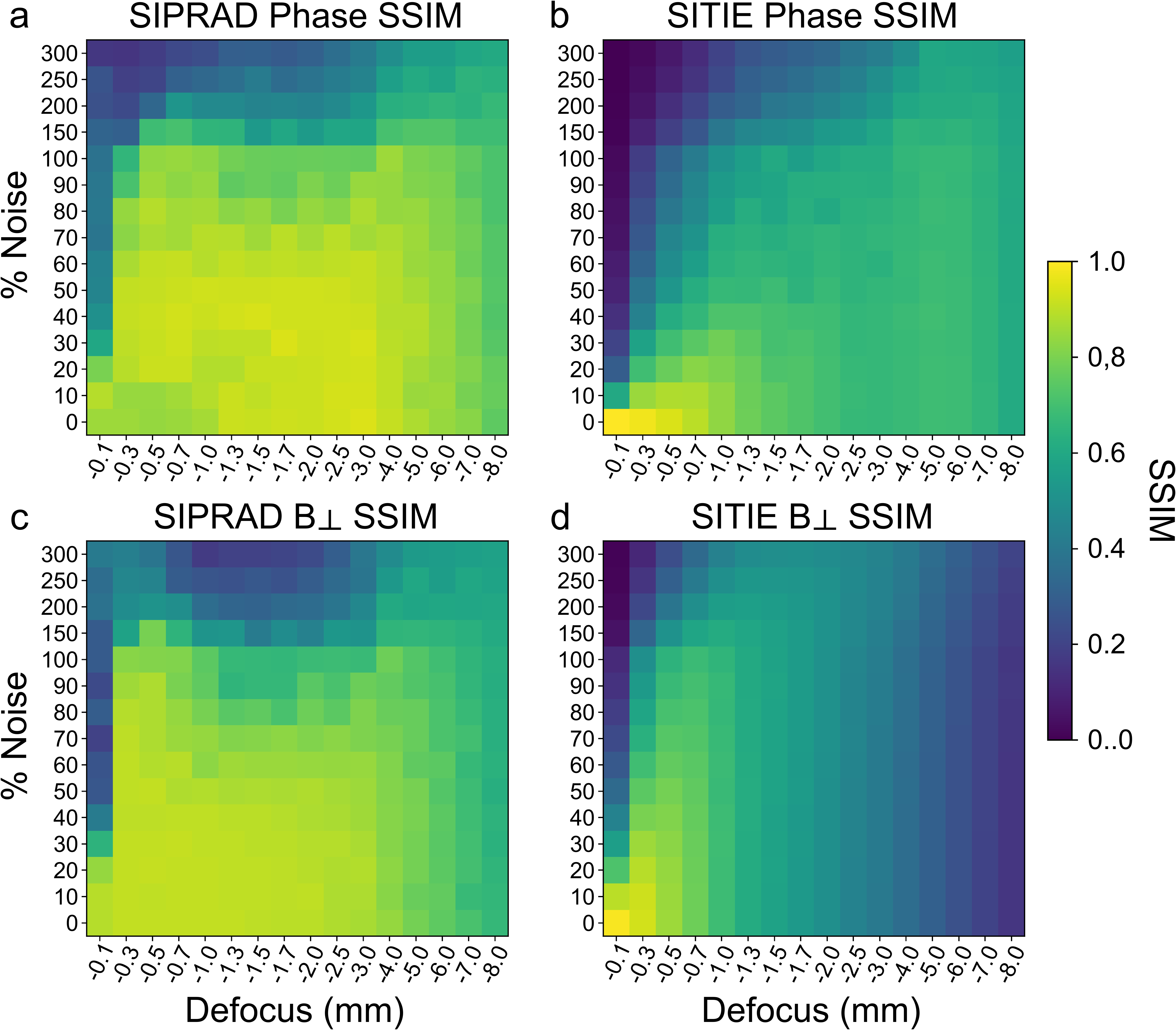}
  \caption{\textbf{Comparison of SSIM for SIPRAD and SITIE reconstructions. a, b} SIPRAD and SITIE phase SSIM plotted on a color scale as a function of \% noise and defocus of the input image. \textbf{c, d} Same as \textbf{(a)} and \textbf{(b)} but plotting SSIM of the integrated induction on the same color scale.}
  \label{figSup:SSIM}
\end{figure}

\clearpage 
\section{Experimental phase reconstruction with a uniform amplitude map}
In Figure 5 of the main text, the SIPRAD reconstruction of the experimental data was performed using a fixed but non-uniform amplitude map obtained by thresholding the input LTEM image. This allowed the effective separation of the magnetic phase shift $\phi_m$ from the total phase shift $\phi$ and the effects of contaminates on the sample. Performing a SIPRAD reconstruction with a fixed and uniform amplitude map leads to a worse phase reconstruction as we show here. The amplitude and $\phi_e$ effects are reflected in the final phase reconstruction and lead to inaccurate features in the integrated induction map. 

\begin{figure}[htb]
\centering
  \includegraphics[width=1.0 \linewidth]{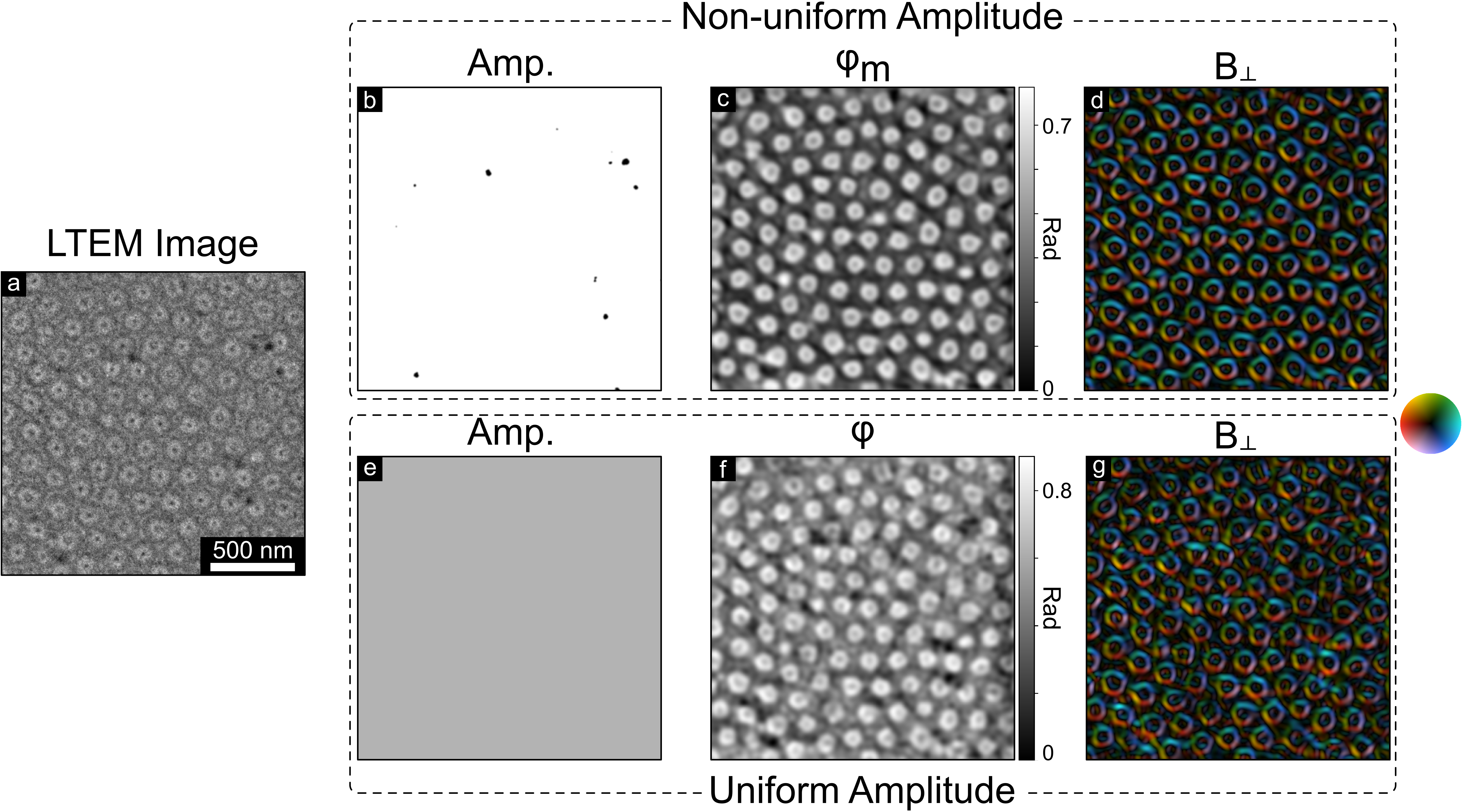}
  \caption{\textbf{SIPRAD phase reconstruction of experimental data with a uniform and non-uniform amplitude map. a} Experimental input LTEM image of CGT. \textbf{b-d} Fixed amplitude and resulting SIPRAD reconstruction as shown in main text Figure 5. \textbf{e} Fixed and uniform amplitude map used for SIPRAD reconstruction. \textbf{f, g} SIPRAD reconstructed phase shift and integrated induction map using the fixed and uniform amplitude.}
  \label{figSup:expNoAmp}
\end{figure}

\clearpage
\section{Through-focal series of CGT}
Here we show the through-focal series (TFS) of \ce{Cr2Ge2Te6} (CGT) used for the TIE reconstruction in Figure 5 of the main text. These images are acquired at \qty{23}{\kelvin}, $\Delta z = \pm \qty{700}{\micro\meter}$, and with a \qty{500}{\Oe} applied out-of-plane field. The dark contrast that remains in the in-focus image is due to non-magnetic surface contaminates that impart both an electrostatic phase shift and create a non-uniform amplitude of the exit electron wave. The underfocus image only is used for SIPRAD and SITIE reconstructions.

\begin{figure}[htb]
\centering
  \includegraphics[width=0.9 \linewidth]{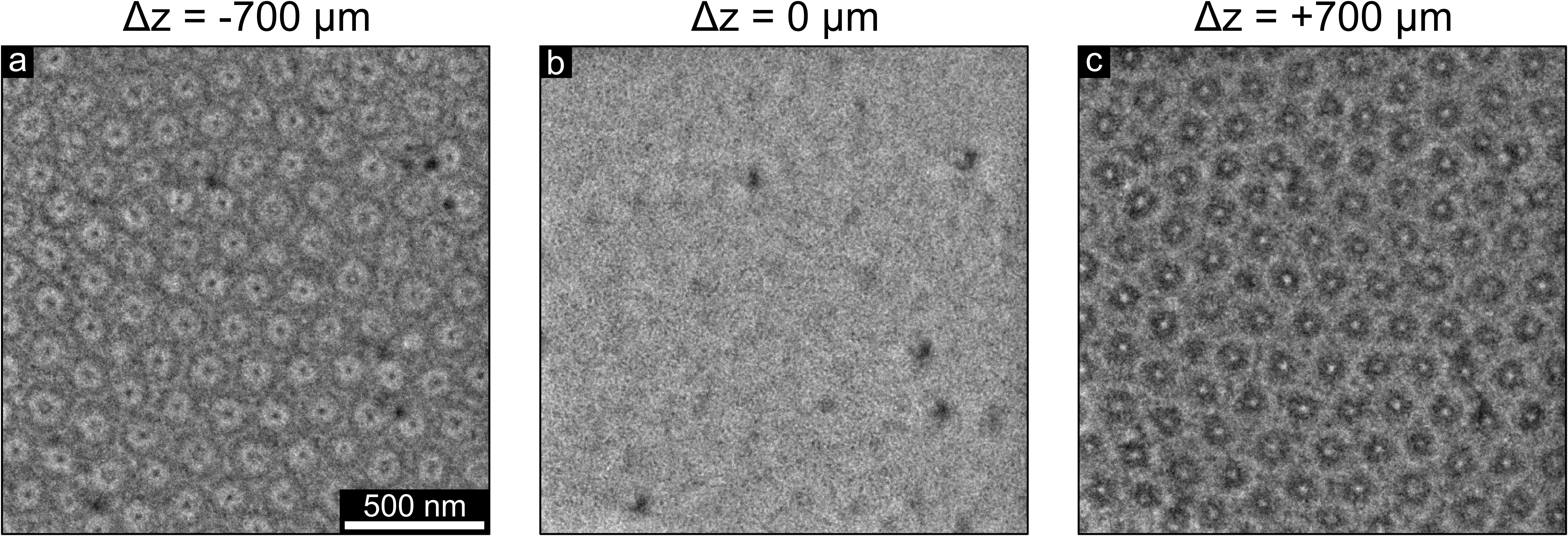}
  \caption{\textbf{Experimental TFS of CGT. a} Under- \textbf{b} in- and \textbf{c} over-focus images of CGT.}
  \label{figSup:TFS}
\end{figure}

\clearpage
\section{Line plots of experimental induction maps}
Here we show line plots of the integrated induction maps reconstructed from experimental data in main text Figure 5. For SIPRAD, SITIE, and TIE we plot the $y$ component of the integrated magnetic induction across the same bubble. We compare that to the true integrated induction of a simulated magnetic bubble. While none of the experimental data perfectly matches the simulation, the SIPRAD reconstructed induction most accurately corresponds to the true domain wall width and intensity. 

\begin{figure}[htb]
\centering
  \includegraphics[width=0.9 \linewidth]{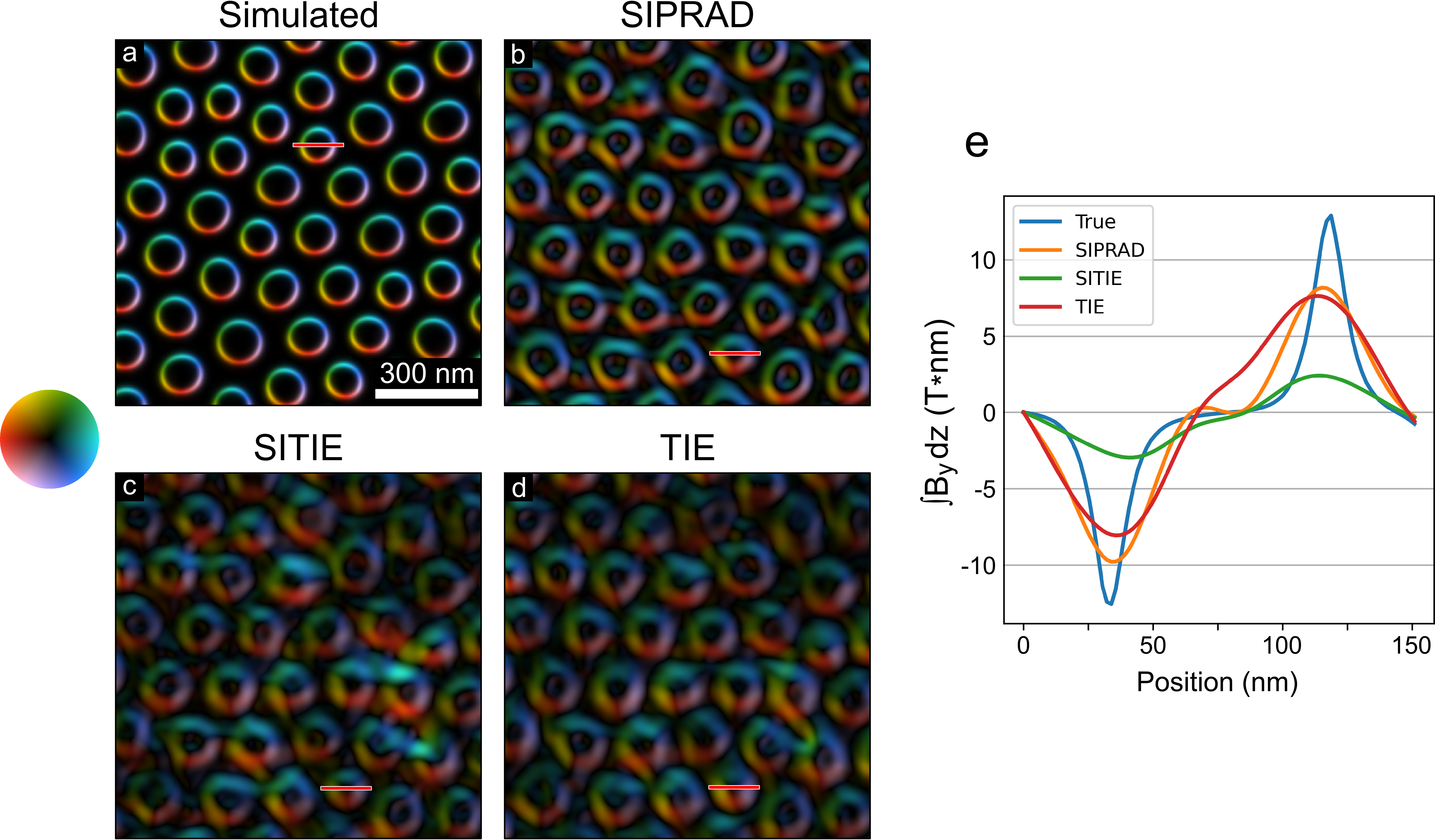}
  \caption{\textbf{Comparison of experimental integrated induction maps to simulated data. a} Simulated integrated induction map for magnetic bubbles in CGT. \textbf{b-d} Cropped regions of the SIPRAD, SITIE, and TIE integrated induction maps shown in Figure 5 of the main text. \textbf{e} Line plots of the $y$ component of the integrated induction, taken along the red lines in \textbf{a-d}.}
  \label{figSup:expLines}
\end{figure}

\clearpage
\section{Deep image prior architecture}

\begin{figure}[htb]
\centering
  \includegraphics[width=0.7 \linewidth]{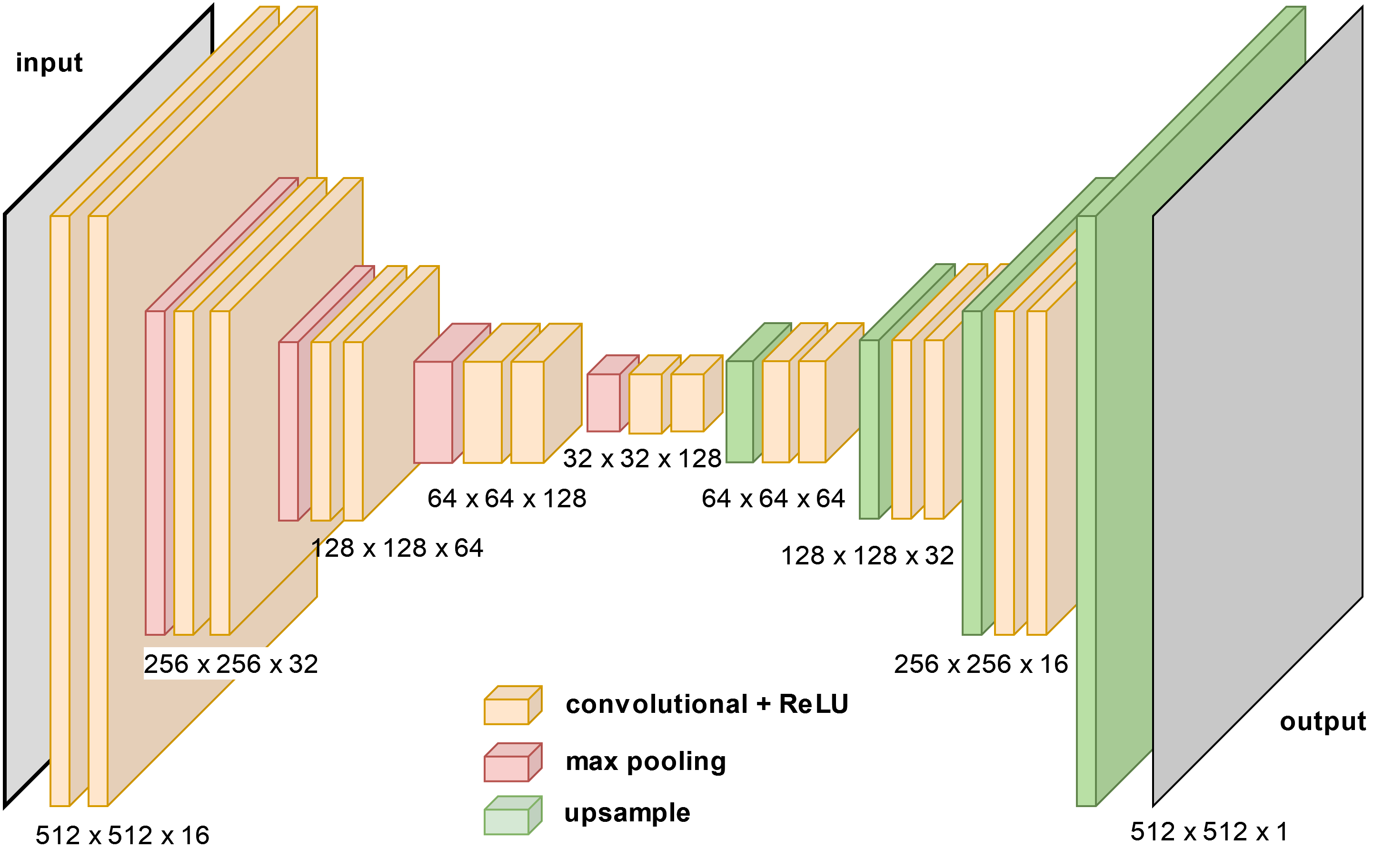}
  \caption{Architecture of the convolutional autoencoders used as the deep image priors in the SIPRAD algorithm.}
  \label{figSup:DIP}
\end{figure}
